\documentclass{JHEP3}
\usepackage{graphicx,amsmath}
\newcommand{\alt}{\mathrel{\raisebox{-.6ex}{$\stackrel{\textstyle<}{\sim}$}}}
\newcommand{\agt}{\mathrel{\raisebox{-.6ex}{$\stackrel{\textstyle>}{\sim}$}}}
\newcommand{\beq}     {\begin{equation}}
\newcommand{\eeq}     {\end{equation}}
\newcommand{\bea}     {\begin{eqnarray}}
\newcommand{\eea}     {\end{eqnarray}}

\newcommand{\stb}     { s_{2\beta}}
\newcommand{\cw}     { c_W}
\newcommand{\sw}     { s_W}
\newcommand{\no}      {\nonumber}

\newcommand{\es}      {\epsilon}

\newcommand{\Sg}      {\Sigma}

\newcommand{\dt}      {\delta}

\newcommand{\mz}      {m_Z}

\newcommand{\MN}      {\mathcal{M}_N}
\newcommand{\MC}      {\mathcal{M}_C}
\def\neu{\widetilde{\chi}^0}
\def\cha{\widetilde{\chi}^\pm}

\title{A minimal supersymmetric scenario with only $\mu$ at the weak scale}

\author{Kingman Cheung \\
Department of Physics and NCTS, National Tsing Hua University,
Hsinchu, Taiwan, R.O.C. \\
Email: {\tt cheung@phys.nthu.edu.tw} }

\author{Cheng-Wei Chiang \\
Department of Physics, National Central University, Chungli, Taiwan 320,
R.O.C. \\
Institute of Physics, Academia Sinica, Taipei, Taiwan 115, R.O.C. \\
Email: {\tt chengwei@phy.ncu.edu.tw} }

\author{Jeonghyeon Song \\
Department of Physics, Konkuk University, Seoul 143-701, Korea \\
Email: {\tt jhsong@konkuk.ac.kr} }

\preprint{hep-ph/0512192}
\abstract{
  Inspired by split supersymmetry, we study a minimal supersymmetric scenario
  with only the Higgsino mass parameter $\mu$ below the TeV scale.  The
  motivation is to satisfy the gauge coupling unification and dark matter
  constraints with the minimal particle contents at the electroweak scale in
  supersymmetric models.  With the neutral Higgsino as the lightest
  supersymmetric particle, we discuss the dark matter signals in both direct
  and indirect detection.  We also discuss collider phenomenology associated
  with the two lightest neutralinos and the lightest chargino, which are almost
  degenerate in mass even after taking into account radiative corrections.
  Unfortunately, the collider signals may be very difficult for identifying
  such a scenario because the pions or leptons in the final state are too soft.
}
\keywords{Supersymmetry, Dark matter, Collider Phenomenology, Minimal}
\begin{document}

\section{Introduction}
\label{sec:introduction}

Supersymmetry (SUSY) is one of the leading candidates for physics beyond the
standard model (SM).  Many of its virtues, including solving the gauge
hierarchy problem, gauge coupling unification, dynamical electroweak symmetry
breaking, and providing a dark matter (DM) candidate, make it one of the most
studied theories.  Nevertheless, generic supersymmetric models suffer from
various problems such as unsuppressed flavor-changing neutral currents (FCNC),
many CP-violating phases, which potentially give rise to large electric dipole
moments (EDM's) as well as other CP-violating phenomena, and too many soft SUSY
breaking parameters.  In order to satisfy the experimental constraints on FCNC
and CP violation, the masses of the scalar fermions are pushed to
multi-TeV~\cite{tev}, or the CP phases are set at extremely small values or
fine-tuned to cancel each other~\cite{nath}.  In particular, if one pushes the
first possibility even further so that the scalar fermion masses are extremely
large, CP violation and FCNC problems no longer exist.  Of course, this
re-introduces the fine-tuning problem, and thus one of the motivations for SUSY
is lost.

From the landscape point of view in string theories with a huge number of
vacua, however, it is not impossible and even more likely to find a vacuum with
a high SUSY breaking scale.  Based on this observation, Arkani-Hamed and
Dimopoulos adopted a rather radical approach to SUSY
breaking~\cite{arkani,kachru}, which was later coined as split
SUSY~\cite{giudice}.  They essentially discarded the hierarchy problem by
accepting the fine-tuning solution to the Higgs boson mass.  All the scalars,
except for a CP-even Higgs boson, are very heavy.  A common mass scale is
usually assumed at $\tilde{m} \sim 10^{9}$ GeV to $M_{\rm GUT}$.  However, the
gaugino masses $M_i$ and the Higgsino mass parameter $\mu$ are comparatively
light at the TeV scale, in order to provide an acceptable dark matter
candidate~\cite{wmap} and to ensure gauge coupling unification.

Previously, two of us proposed a further splitting in split SUSY by raising the
$\mu$ parameter to a large value which could be about the same as the sfermion
mass or the SUSY breaking scale~\cite{cw}.  It was called the high-$\mu$ split
SUSY scenario.  In such a scenario, we do not encounter the notorious $\mu$
problem~\cite{mu}, a viable dark matter candidate is still available, and the
gauge coupling unification is only slightly worsened.

In this work, we study a complementary scenario in which only the Higgsino mass
parameter $\mu$ remains at the weak scale while all other soft SUSY breaking
parameters are pushed to $\tilde{m}\;(10^9\;{\rm GeV})$.  In the framework of
minimal supersymmetric standard model (MSSM), this is a minimal scenario since
only the SM particle masses and the Higgsino mass parameter $\mu$ remain below
the weak or TeV scale.  The spectrum is very simple: the SM particles, a light
CP-even Higgs boson, two neutral Higgsinos and a pair of charged Higgsinos.
All the other SUSY particles are at the very high SUSY breaking scale.  It can
be called the low-$\mu$ split SUSY.  Although it may be difficult to generate
such a scenario from a sensible SUSY breaking model \footnote{Although some
  models~\cite{wells,model} are constructed to give large hierarchies among
  soft parameters, it is hard to give more than a few orders of magnitude
  difference.}, its phenomenology is so special and unique at colliders that
the model deserves a good study, which is the primary goal of the paper.
Note that the large separation of scales between the $\mu$ parameter
and the gaugino masses will induce a radiative correction to the $\mu$
parameter such that the $\mu$ parameter is within a loop factor of the
gaugino mass.  We therefore need another fine-tuning between the correction
and the bare parameter so that the physical $\mu$ parameter is at the
electroweak scale while the gaugino masses stay at a high scale.

We summarize the differences between low-$\mu$ split SUSY and ordinary split
SUSY as follows:
\begin{enumerate}
\item The gaugino masses are raised to a very high scale in this scenario while
  in ordinary split SUSY they are kept at the electroweak scale.
\item The lightest supersymmetric particle (LSP) is the neutral Higgsino,
  contrary to the ordinary split SUSY model where the Bino, wino, and Higgsino
  are all possible to be the LSP as the dark matter candidate
  \cite{aaron,stefano,Wang:2005kf}
\item In the low-$\mu$ split SUSY scenario, the Higgsino dark matter has a
  negligible elastic scattering cross section with the nuclei in direct
  detection methods.  Instead the Higgsino pair annihilation into gauge boson
  pairs, particularly a pair of monochromatic photons, is strong.  The Bino
  dark matter in ordinary split SUSY has only a small pair annihilation rate
  into gauge bosons and diphotons.
\item In low-$\mu$ split SUSY, the two lightest neutralinos and the lightest
  chargino have almost degenerate masses.  Even the radiative corrections
  cannot lead to a mass difference more than one GeV between the lightest
  chargino and the lightest neutralino.  Therefore, the decay of the lightest
  chargino only produces very soft pions or leptons in the final state, which
  may be too difficult to detect at high energy colliders.  In split SUSY, the
  chargino and neutralino can give rise to interesting signals at hadron and
  $e^+ e^-$ colliders~\cite{kilian,zhu,song}
\end{enumerate}

Note that the present low-$\mu$ split SUSY scenario is rather similar to the
focus-point supersymmetry in the phenomenological aspects~\cite{feng}.
Recently, there have been works exploiting the idea of minimal extensions of
the SM to satisfy the dark matter and other constraints \cite{mura,marco}.

The organization of the paper is as follows.  In the next section, we examine
the gauge coupling unification.  In Sec.~\ref{sec:sp}, we describe the mass
spectrum of the neutralinos and charginos.  Section \ref{sec:rel-coupling}
deals with the couplings relevant to the studies of dark matter and collider
phenomenology.  In Sec.~\ref{sec:dm}, we discuss the dark matter relic density
and the direct and indirect detection.  In Sec.~\ref{sec:pheno}, we study the
phenomenology at hadron and $e^+ e^-$ colliders.  We conclude in Sec.
\ref{sec:summary}.

\section{Gauge Coupling Unification
\label{sec:unification}}

\begin{figure}[th!]
\centering
\includegraphics[width=4in]{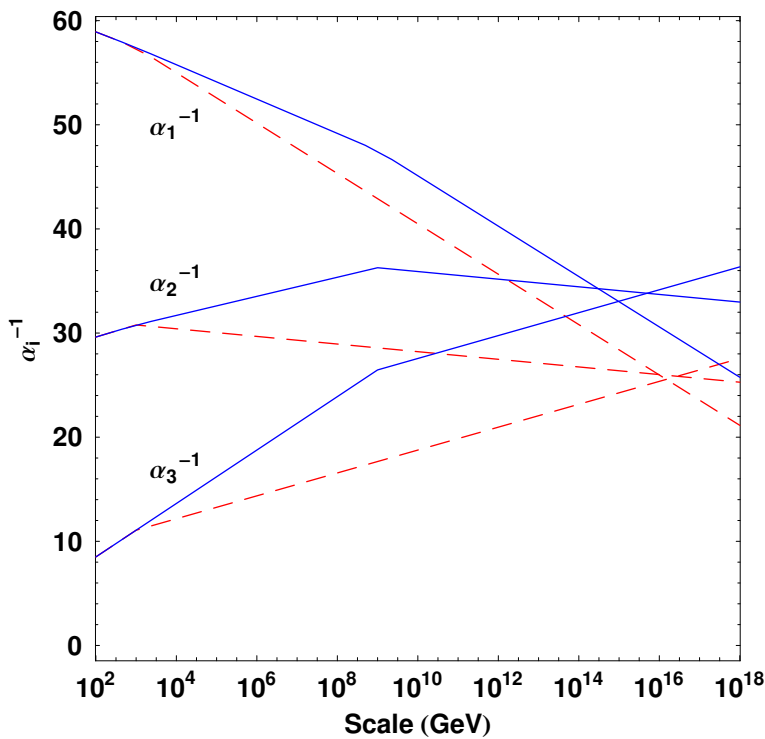}
\caption{\small \label{unif}
  Gauge coupling unification at the one-loop level.  In the low-$\mu$ split
  SUSY scenario (indicated by solid curves), the Higgsino masses are set at 1
  TeV while all the other soft SUSY parameters at $10^9$ GeV.  The gauge
  coupling running in the MSSM (dashed curves) is also included for comparison.
  Following Ref.~\cite{arkani}, we take $\alpha_1^{-1}(M_Z) = 58.98$,
  $\alpha_2^{-1}(M_Z) = 29.57$, and $\alpha_3^{-1}(M_Z) = 8.40$.}
\end{figure}

The general form of the one-loop renormalization group equations for the gauge
couplings between any two mass scales $M_X$ and $M_Y$ is given by
\begin{equation}
\frac{1}{\alpha_i(M_X^2)} = \frac{1}{\alpha_i ( M_Y^2)} -
 \frac{\beta_i}{4 \pi} \ln \left( \frac{M_X^2}{M_Y^2} \right ) ~,
\end{equation}
where $i = 1,2,3$ are indices representing the $U(1)_Y$, $SU(2)_L$, and
$SU(3)_C$ gauge couplings, respectively.  The differences among the SM, MSSM,
ordinary split SUSY, and low-$\mu$ split SUSY scenarios reside in the following
values of the beta functions :
\begin{eqnarray}
\mbox{SM}: && (\beta)_{\rm SM} = \left( \begin{array}{c}
                   0 \\
                  - \frac{22}{3} \\
                  - 11 \end{array}         \right )
           +  \left( \begin{array}{c}
                   \frac{4}{3} \\
                   \frac{4}{3} \\
                   \frac{4}{3}    \end{array} \right ) F
           +  \left( \begin{array}{c}
                   \frac{1}{10} \\
                   \frac{1}{6} \\
                   0            \end{array} \right ) N_H ~,
  \nonumber \\
\mbox{MSSM}: &&  (\beta)_{\rm MSSM} = \left( \begin{array}{c}
                    0 \\
                  - 6 \\
                  - 9 \end{array}         \right )
           +  \left( \begin{array}{c}
                   2 \\
                   2 \\
                   2    \end{array} \right ) F
           +  \left( \begin{array}{c}
                   \frac{3}{10} \\
                   \frac{1}{2} \\
                   0            \end{array} \right ) N_H ~,
  \nonumber \\
\mbox{Split-SUSY}: && (\beta)_{\rm split}|_{< \tilde{m}}= \left(
  \begin{array}{c}
                    0 \\
                  - 6 \\
                  - 9 \end{array}         \right )
           +  \left( \begin{array}{c}
                   \frac{4}{3} \\
                   \frac{4}{3} \\
                   \frac{4}{3}    \end{array} \right ) F
           +  \left( \begin{array}{c}
                   \frac{5}{10} \\
                   \frac{5}{6} \\
                   0            \end{array} \right ) ~,
  \nonumber \\
\mbox{low-$\mu$ split SUSY}: &&
(\beta)_{\mu{\rm -split}}|_{< \tilde{m}}= \left( \begin{array}{c}
                    0 \\
                  - 22/3 \\
                  - 11 \end{array}         \right )
           +  \left( \begin{array}{c}
                   \frac{4}{3} \\
                   \frac{4}{3} \\
                   \frac{4}{3}    \end{array} \right ) F
           +  \left( \begin{array}{c}
                   \frac{5}{10} \\
                   \frac{5}{6} \\
                   0            \end{array} \right ) ~,
  \nonumber
\end{eqnarray}
where $F=3$ is the number of generations of fermions or sfermions, and $N_H$ is
the number of Higgs doublets ($N_H=1$ in the SM, $N_H=2$ in the SUSY).  In the
evolution of the gauge couplings in the low-$\mu$ split SUSY scenario, we use
(i) the SM $\beta_i$'s from the weak scale ($\mz$) to $\mu$, the scale of
Higgsino masses, which we take a common value of 1 TeV;
(ii) the $\beta_i$'s in low-$\mu$ split SUSY scenario from $\mu$ to
$\tilde{m}$, which we fix at $10^9$ GeV; and
(iii) the MSSM $\beta_i$'s from $\tilde{m}$ to the grand unified scale.

The unification of gauge couplings in the low-$\mu$ split SUSY case is
slightly worse than that in the MSSM, but significantly better than
that in the SM.  The unification scale is about $10^{14.5}$ GeV. Such
a low-scale unification can be dangerous in standard GUTs if there is
no other additional symmetries or mechanisms to protect the proton
from decaying. One solution is to have a higher dimensional orbifold
GUT with a reduced gauge symmetry on the boundary
\cite{Hall:2002ea}. This also solves the doublet-triplet splitting
problem in the usual grand unified models. In this case, threshold
effects due to Kaluza-Klein modes have to be included in the running
between the compactification scale and the grand unified scale.

If we take the intersection of the $\alpha_1$ and $\alpha_2$ curves as the
value of gauge coupling strength $\alpha_{\rm GUT}$ at the unification scale
(here $\alpha_{\rm GUT} \simeq 1/34.3$, smaller than that in the MSSM) and run
it down to the $\mz$ scale for the strong coupling, we obtain $\alpha_3(\mz) =
0.098$.  For comparison, the measured value is $\alpha_3(\mz)_{\rm exp} =
0.1182 \pm 0.0027$~\cite{Bethke:2004uy} and the one-loop prediction in the MSSM
is $\alpha_3(\mz) = 0.110$.  It is well-known that the two-loop effects in the
MSSM increase $\alpha_3(\mz)$ to $0.130$.  Therefore, we expect that the
two-loop contribution in our scenario will lift the predicted $\alpha_3(\mz)$
closer to the observed value.

\section{Mass Spectrum \label{sec:sp}}

In the low-$\mu$ split SUSY scenario, the approximation of $\mu \ll M_{1}< M_2$
is very effective.  Then the tree-level neutralino masses can be expanded in
terms of $\delta= \mu/M_1$:
\begin{eqnarray}
m_{\neu_1}' &=& \mu
\left[1-\frac{1}{2}(1-\stb)(\cw^2 + r_2 \sw^2) \frac{r_z^2}{r_2} \dt +
\mathcal{O}(\dt^2)\right], \\
m_{\neu_2}' &=& -\mu
\left[1+\frac{1}{2}(1+\stb)(\cw^2 + r_2 \sw^2) \frac{r_z^2}{r_2} \dt +
\mathcal{O}(\dt^2)\right], \\
m_{\neu_3}' &=& M_1 \left[
1+\mathcal{O}(\dt^2)
\right],
\\
m_{\neu_4}' &=& M_2 \left[
1+\mathcal{O}(\dt^2)
\right]
\,,
\end{eqnarray}
where $s_W = \sin\theta_{\rm w}$, $c_W=\cos\theta_{\rm w}$, $s_{2\beta} = \sin
2\beta$, $r_2 = M_2/M_1$, and $r_z=m_Z/\mu$.  $\theta_{\rm w}$ is the Weinberg
angle, and $\tan\beta$ is the ratio of the two vacuum expectation values in the
Higgs sector.  The diagonalizing matrix $N$~\cite{Haber:1984rc} in the leading
order becomes
\begin{eqnarray}
\label{eq:N-leading}
N &=& \left(\begin{array}{cccc}
          0 & 0 & \frac{1}{\sqrt{2}} & \mp \frac{1}{\sqrt{2}} \\
          0 & 0 & \pm\frac{1}{\sqrt{2}} & \frac{1}{\sqrt{2}}\\
          1 & 0 & 0 & 0\\
          0 & 1 & 0 & 0
        \end{array}
        \right)\;, \qquad \qquad {\rm for} \;\;
\mu\; {}^{>}_{<} \; 0.
\end{eqnarray}
The chargino masses are
\beq
m_{\cha_1} = \left| \mu \left(1+\stb
\frac{m_W^2}{M_2 \mu} \right) \right|,
\quad m_{\cha_2} = M_2
\left(1+\stb \frac{m_W^2}{M_2^2}\right),
\eeq
which are obtained by the diagonalizing matrices $U$ and
$V$~\cite{Haber:1984rc} given as, in the leading order,
\begin{eqnarray}
\label{eq:UV-leading}
U &=&
\left(\begin{array}{cc}
        0 & \pm 1 \\
        1 & 0
      \end{array}
\right)\;, \qquad
V =
\left(\begin{array}{cc}
        0 & 1 \\
        1 & 0
      \end{array}
\right) \quad \quad \hbox{for } \mu \; {}^{>}_{<}\;0 \;.
\end{eqnarray}

At tree level, the mass differences with respect to the lightest neutralino are
\begin{eqnarray}
\Delta M^{\rm tree}_{12}
&\equiv & m_{\widetilde{\chi}^0_2} - m_{\widetilde{\chi}^0_1}
=
\left[
1 + t_W^2 \frac{M_2}{M_1}
\right]
\frac{m_W^2}{M_2}, \\
\Delta M^{\rm tree}_{+}
&\equiv & m_{\widetilde{\chi}^+_1}-m_{\widetilde{\chi}^0_1}
= \frac{1}{2}
\left[
1+(1-s_{2\beta})\,
        t_W^2  \frac{M_2}{M_1}
\right]
\frac{m_W^2}{M_2}  \;,
\end{eqnarray}
where $t_W=s_W/c_W$.  It is easy to see that if $\tan\beta \gg 1$, $\Delta
M^{\rm tree}_{12} \simeq 2 \,\Delta M^{\rm tree}_{+}$.  For $M_1 \sim M_2 \sim
10^8 - 10^9$ GeV, $\Delta M^{\rm tree}_{12} \simeq 2 \,\Delta M^{\rm tree}_{+}
\sim \mathcal{O}(10-100)$ keV.

With heavy gaugino masses at the order of $10^9$ GeV, the mass differences due
to radiative corrections are more important than the tree-level mass
differences.  The one-loop radiative mass corrections to neutralinos and
charginos were performed in Ref.~\cite{Pierce}, where the complete one-loop
self-energies for charginos and neutralinos are included in their mass
matrices.  The total mass matrices are diagonalized to obtain the final mass
spectrum.  The one-loop neutralino mass matrix is
\beq
\MN = \MN ^{(0)} +
\frac{1}{2} \left[
\dt \MN (p^2) + \dt \MN^T (p^2)
\right]
\,,
\eeq
and the one-loop chargino mass matrix is
\beq
\MC = \MC^{(0)} -\Sg^+_R (p^2) \MC -\MC \Sg_L^+(p^2) -\Sg_S^+ (p^2).
\eeq
The detailed expressions for $\MN$ and $\MC$ are referred to
Ref.~\cite{Pierce}.  We note that the one-loop neutralino mass matrix is still
symmetric.

The mass degeneracy between $\cha_1$ and $\neu_1$ is lifted by the radiative
corrections, given by
\beq \Delta M^{\rm rad}_+
=
\frac{g^2}{8 \pi^2} \sw^2 \mu
\left[
B_1(\mu,\mu,\mz)-2 B_0(\mu,\mu,\mz) - \left\{
B_1(\mu,\mu,0)-2 B_0(\mu,\mu,0)
\right\}
\right]
\,,
\eeq
where we have neglected the contributions from super heavy particles like
sfermions and non-SM Higgs bosons.  Using the relation
\beq
2\big[B_1(M,M,m)-2 B_0(M,M,m)\big] = \frac{1}{\es}+f\left(\frac{m}{M} \right),
\eeq
the mass difference is simplified to
\beq
\Delta M^{\rm rad}_+
=
\frac{\alpha_2 \mu }{4 \pi} \sw^2 \left[
f\left(\frac{\mz}{\mu} \right)
- f(0)
\right]
\,.
\eeq
The function $f(a)$ is \cite{nojiri}
\beq
f(a) = \int_0^1 d x\, 2 (1+x)
\ln(x^2+(1-x) a^2) \simeq -5 + c\, a + {\mathcal O}(a^2),
\eeq
in which the second equality holds for $a \ll 1$.  Numerically, $c$ is about
6.3.  For $\mu \gg m_{Z}$, we have $\Delta M^{\rm rad}_+$ almost independent of
$\mu$, given by
\beq
\Delta M^{\rm rad}_+ = \frac{\alpha_2}{4 \pi } c \, \mz \,\sw^2
  \quad \hbox{for }
\mu \gg \mz
\,.
\eeq
Numerically it is about $340$ MeV, much larger than the tree-level splitting
for $M_{1,2} \simeq 10^9$ GeV.  For comparison, we present the radiative mass
difference in the wino-LSP scenario~\cite{nojiri}:
\beq
\Delta M_+^{\rm rad} |_{\rm wino} 
=\frac{\alpha_2}{4 \pi } c \, \mz \,\cw(1-\cw) \quad \hbox{for }
\mu \gg \mz.
\eeq
Since $\cw(1-\cw)\simeq 0.11$ for $s^2_W \simeq 0.23$, the radiative mass
correction for the Higgsino case is about twice as much as the wino case.
Therefore, the lightest chargino will decay into the lightest neutralino plus a
soft pion, or a charged lepton with a neutrino.  We cannot avoid the decay of
the lightest chargino into a pion.

On the other hand, $\Delta M_{12}(\equiv m_{\neu_2}-m_{\neu_1})$ remains intact
from the radiative mass corrections in the low-$\mu$ split SUSY case.  The
radiative corrections modify the neutralino mass matrix in the way preserving
the symmetry property, $\MN=\MN^T$.  Neglecting the extremely small $\mu/M_1$
correction, $\Delta M_{12}^{\rm rad}$ comes from the difference between the
$(\MN)_{34}$ and $(\MN)_{43}$ components, which are the same.  $\neu_1$ and
$\neu_2$ are, therefore, highly degenerate with a mass difference of the order
of $(10-100)$ keV in the low-$\mu$ split SUSY scenario.  This small difference
prohibits the decay channel of $\neu_2 \to \neu_1 Z^* \to \neu_1 e^+ e^-$.  The
only open modes are $\neu_2 \to \neu_1 Z^* \to \neu_1 \nu \bar{\nu}$ and
$\neu_2 \to \neu_1 \gamma$.  Unfortunately both are difficult to probe, as the
former will be purely the missing energy signal and the latter will generate
too soft photons with missing energy.

\section{Relevant Couplings}
\label{sec:rel-coupling}

As the gaugino mass parameters $M_1$, $M_2$ and $M_3$ become very large while
the $\mu$ parameter remains light, only the two lightest neutralinos and the
lightest chargino are accessible at high energy colliders.  In this section, we
highlight the couplings that are relevant to studies on collider and dark
matter phenomenology.  Let us first examine their relevant couplings to gauge
bosons and the Higgs bosons.

\begin{itemize}
\item The $Z$-$\widetilde{\chi}^0_{1,2}$-$\widetilde{\chi}^0_{1,2}$ couplings
  only receive contributions from the Higgsino-Higgsino-gauge couplings.  In
  the limit of very large $M_1$ and $M_2$ the Higgsino component of
  $\widetilde{\chi}^0_1$ and $\widetilde{\chi}^0_2$ are essentially one.
  Therefore, the $Z$ boson couplings to the neutralinos are large.
\item The $H$-$\widetilde{\chi}^0_{1,2}$-$\widetilde{\chi}^0_{1,2}$ couplings
  have sources from the Higgs-Higgsino-gaugino terms.  Therefore, in the limit
  of large $M_1$ and $M_2$, these couplings go to zero.
\item The $W^-$-$\widetilde{\chi}^0_{1,2}$-$\widetilde{\chi}^+_1$ couplings
  have sources from the Higgsino-Higgsino-gauge couplings and from the
  gaugino-gaugino-gauge couplings.  In the limit of large $M_1$ and $M_2$, the
  latter contribution goes to zero while the former remains.  Therefore, the
  $W^-$-$\widetilde{\chi}^0_{1,2}$-$\widetilde{\chi}^+_1$ couplings contain
  only the Higgsino-Higgsino-gauge part.
\item The $H^-$-$\widetilde{\chi}^0_{1,2}$-$\widetilde{\chi}^+_1$ couplings
  have sources from the Higgs-Higgsino-gaugino couplings. In the limit of large
  $M_1$ and $M_2$, they do not contribute to
  $H^-$-$\widetilde{\chi}^0_{1,2}$-$\widetilde{\chi}^+_1$, which thus vanish.
\item The $\gamma(Z)$-$\widetilde{\chi}^+_{1}$-$\widetilde{\chi}^-_1$ couplings
  have sources from the Higgsino-Higgsino-gauge couplings and from the
  gaugino-gaugino-gauge couplings.  In the limit of large $M_1$ and $M_2$, the
  latter contribution goes to zero while the former remains.  Therefore, the
  $\gamma(Z)$-$\widetilde{\chi}^+_{1}$-$\widetilde{\chi}^-_1$ coupling contains
  only the Higgsino-Higgsino-gauge part.
\item The $H$-$\widetilde{\chi}^+_{1}$-$\widetilde{\chi}^-_1$ couplings have
  sources from the Higgs-Higgsino-gaugino couplings. In the limit of large
  $M_1$ and $M_2$, they do not contribute to
  $H$-$\widetilde{\chi}^+_{1}$-$\widetilde{\chi}^-_1$, which thus vanishes.
\end{itemize}

The only couplings that survive in the low-$\mu$ split SUSY scenario are
$Z$-$\widetilde{\chi}^0_{1,2}$-$\widetilde{\chi}^0_{1,2}$,
$W^-$-$\widetilde{\chi}^0_{1,2}$-$\widetilde{\chi}^+_1$, and
$\gamma(Z)$-$\widetilde{\chi}^+_{1}$-$\widetilde{\chi}^-_1$.  The phenomenology
of the two light neutralinos and the lightest chargino depends on the
interaction Lagrangian
\begin{eqnarray}
{\cal L} &=& \frac{g}{2 \cos\theta_{\rm w}} \overline{\widetilde{\chi}^0_i}
\gamma^\mu \,\left( O^{L''}_{ij} P_L + O^{R''}_{ij} P_R \right ) \,
\widetilde{\chi}^0_j \; Z_\mu  \nonumber \\
&+&
\left[ g  \overline{\widetilde{\chi}^0_i}
\gamma^\mu \,\left( O^{L}_{ij} P_L + O^{R}_{ij} P_R \right ) \,
\widetilde{\chi}^+_j \; W^-_\mu   + H.c. \right ]\nonumber \\
&-&
e   \overline{\widetilde{\chi}^+_i} \gamma^\mu
\widetilde{\chi}^+_i \; A_\mu
-
\frac{g}{\cos\theta_{\rm w}} \overline{\widetilde{\chi}^+_i}
\gamma^\mu \,\left( O^{L'}_{ij} P_L + O^{R'}_{ij} P_R \right ) \,
\widetilde{\chi}^+_j \; Z_\mu  \;,
\end{eqnarray}
where
\begin{eqnarray}
O^{L''}_{ij} = \frac{1}{2} \left( N_{i4} N^*_{j4} - N_{i3} N^*_{j3} \right )\;,
&&
O^{R''}_{ij} = - \left(O^{L''}_{ij} \right )^* \;, \nonumber \\
O^{L}_{ij} = N_{i2} V^*_{j1} - \frac{1}{\sqrt{2}} N_{i4} V^*_{j2} \;,
&&
O^{R}_{ij} = N^*_{i2} U_{j1} + \frac{1}{\sqrt{2}} N^*_{i4} U_{j2} \;, \nonumber
\\
O^{L'}_{ij} = \cos^2\theta_{\rm w} V_{i1} V^*_{j1}
  + \frac{\cos2 \theta_{\rm w}}{2} V_{i2} V^*_{j2} \;,
&&
O^{R'}_{ij} = \cos^2\theta_{\rm w} U_{j1} U^*_{i1}
  + \frac{\cos2 \theta_{\rm w}}{2} U_{j2} U^*_{i2} \;. \nonumber
\end{eqnarray}
The mixing matrices in Eqs.~(\ref{eq:N-leading}) and (\ref{eq:UV-leading})
further simplify the couplings as
\bea
\label{eq:simple-couplings}
\hbox{for }\mu\; {}_{<}^{>} \; 0, &&
O^{L''}_{11}= O^{L''}_{22} =0, \quad O^{L''}_{12} =\mp \frac{1}{2}, \no\\
 &&
O^L_{11} = \pm \frac{1}{2}, \quad O^R_{11}= - \frac{1}{2},
 \no \\
&&
O^{L'}_{11}= O^{R'}_{11} = \frac{1}{2} \cos 2 \theta_{\rm w}.
\eea

\section{Dark Matter \label{sec:dm}}

\subsection{Relic Density}
The Higgsino LSP is well-known to have large annihilation cross sections into
$ZZ$ and $WW$ pairs and also into $f\bar f$ via $Z$ boson exchange.  In
addition, the lightest neutralino is close in mass with the second lightest
neutralino and the lightest chargino.  Such large annihilation cross sections
and effective co-annihilation reduce significantly the thermal relic density of
the Higgsino LSP.  The simplified formula for the Higgsino LSP is given by
\[
\Omega_{\tilde{H}} h^2 \simeq 0.1 \, \left( \frac{M_{\rm LSP}}{1\,{\rm TeV}}
\right )^2
\,.
\]
Only heavy neutral Higgsino with mass $\sim 1$ TeV can explain the dominant
dark matter if thermal production is the only source.  Unfortunately, in this
case SUSY particles can only be marginally produced at the LHC.

On the other hand, there can be non-thermal sources of dark matter.  In such a
case, the mass of the Higgsino can be lower.  Both collider and dark matter
experiments can find interesting signals.  A good example of non-thermal
sources can be found in anomaly-mediated SUSY breaking models where the neutral
wino is the LSP \cite{anom,anomaly}.  For a relatively light neutral wino it
cannot be the dominant dark matter because of its large annihilation cross
sections.  However, an intriguing source of non-thermal wino for compensation
is the decay of moduli fields \cite{moroi}, which can produce a sufficient
amount of neutral winos.  This case is similar to the Higgsino dark matter.
There are also other non-thermal sources of Higgsino or wino dark matter
discussed in the literatures \cite{nonthermal}, such as the gravitino NLSP
decay and decay of some scalar relics.

These non-thermal sources have important constraints from cosmology; their late
decay should be before the big bang nucleosynthesis (BBN) and the reheating
temperature due to the decay of the moduli should be above MeV, so as not to
disturb the success of the BBN.  The moduli decay when the expansion rate of
the Universe becomes compatible with the decay width of the moduli.  In the
model used in Ref. \cite{moroi}, the reheating temperature being above MeV
requires the mass of the modulus field to be above 100 TeV.  Therefore, as long
as the decay of moduli is before BBN and the reheating temperature is above
MeV, it is consistent with BBN.

We also note that dark matter can be made up of some almost non-interacting
particles, e.g., axions, which have nothing to do with the electroweak scale
physics.  An interesting scenario of this kind \cite{barger} is proposed, in
which all sparticles are super-heavy and the dark matter is explained by the
axion.

\subsection{Direct Dark Matter Detection}

In general, the neutralino LSP interacts with the nucleon of the detecting
material via Higgs boson exchange and/or squark exchange for the
spin-independent cross section while via $Z$ boson exchange and part of the
squark exchange for spin-dependent cross section.  Since the signal is very
mild, a large coherence effect in the detecting material is crucial, which is
only useful for the spin-independent cross section.  In the low-$\mu$ split
SUSY scenario, however, the Higgs-Higgsino-gaugino couplings are zero because
of the decoupling of the gauginos, and the squarks are extremely heavy.  Thus,
the spin-independent cross section is negligible \cite{darksusy}.

Here we concentrate on the spin-dependent cross sections.  Even though the
current detector sensitivity is much lower, there are some proposals that focus
on measuring the spin-dependent cross sections \cite{spin-dep}.  In split SUSY,
the contributions from super-heavy squarks are negligible.  Therefore, the only
remaining contribution comes from the $Z$-boson exchange via the
Higgsino-Higgsino-gauge type coupling.

Let us remind the reader of the spin-dependent cross section with protons and
neutrons:
\begin{equation}
\sigma^{SD}_{\chi p} = \frac{3 \mu_{\chi p}^2}{\pi} \left| G^p_a \right |^2\;,
\qquad
\sigma^{SD}_{\chi n} = \frac{3 \mu_{\chi n}^2}{\pi} \left| G^n_a \right |^2\;,
\end{equation}
where $\mu_{\chi p}$ ($\mu_{\chi n}$) is the reduced mass of the neutralino
and proton (neutron), and the effective axial couplings $G_a^{p,n}$ are
\begin{equation}
G_a^{p,n} = \sum_{u,d,s} \, \left ( \Delta q \right)_{p,n} \,
  \frac{g_{Z\chi \chi} g_{Zqq}}{ m_Z^2} \;,
\end{equation}
where
\[
g_{Z\chi\chi}= \frac{g}{2 \cos\theta_{\rm w}} \left( |N_{13}|^2 - |N_{14}|^2
                                                 \right ) \;, \qquad
g_{Zqq} = - \frac{g}{2 \cos\theta_{\rm w}} T_{3q} \;.
\]
It is easy to see that in this low-$\mu$ split SUSY scenario ($\mu \ll M_1,
M_2$) the $H_u^0$ and $H_d^0$ mix maximally so that $|N_{13}| = |N_{14}| =
1/\sqrt{2}$.  Therefore, the vanishing factor $N_{13}^2 - N^2_{14} \simeq 0$
suppresses the spin-dependent cross section also.

Since the mass difference between $\widetilde{\chi}^0_1$ and 
$\widetilde{\chi}^0_2$ is of the order of 
$ 10 - 100$ keV for $M_2 \sim 10^9 - 10^8$ GeV, there is a slight chance
that the $\widetilde{\chi}^0_1$ transits into $\widetilde{\chi}^0_2$ 
after the elastic scattering with the nuclei in the detector, via
the nonzero $Z$-$\widetilde{\chi}^0_1$-$\widetilde{\chi}^0_2$ coupling.
The velocity of the weakly interacting massive particle (WIMP)
in our galaxy follows a Boltzmann distribution
centered at $v=270$ km s$^{-1} \sim 10^{-3} c$.  The spectrum of recoil
is exponential with a typical energy $\langle E \rangle \sim 50$ keV. 
Therefore, if $M_2 > 10^9$ GeV and so the mass difference 
$\Delta M_{12} \alt $ a few keV, $\widetilde{\chi}^0_1$ can transit 
into $\widetilde{\chi}^0_2$ after the scattering.  
This is because the recoil energy is large enough
for the transition to take place.  
While this could be an interesting signal, we require 
$M_2 \alt 10^9$ GeV so that such transitions cannot take place. 
Moreover, the gluino cosmology requires a SUSY breaking scale to be less
than $10^9$ GeV \cite{arvan}.
Therefore, the direct detection experiments do not have constraints on the
present scenario.

Although the tree-level spin-independent and spin-dependent scattering cross
sections are negligible in this scenario, one-loop corrections give a non-zero
cross section \cite{hisano}.  However, the cross section is only at the
$10^{-45}$ cm$^2$ level \cite{hisano}, which is way below the current and
future experimental limits.

\subsection{Indirect Dark matter detection}

The Higgsino dark matter can have very interesting signals for indirect
detection in view of its large annihilation cross sections into $ZZ$ and $W^+
W^-$ pairs, as well as into $\gamma \gamma, \gamma Z$ via one-loop diagrams.
In particular, the last two channels, though loop-suppressed, can give a very
clean signal of monochromatic photon lines.  If the resolution of the photon
detectors (either ground-based or satellite-based) is high enough, a clean and
unambiguous photon peak at hundreds of GeV can be observed above the
background.

Here we give an estimate of the photon flux in the low-$\mu$ split SUSY
scenario in which only the $W^-$-$\widetilde{\chi}^+_1$ loop is important.
Using the results given in Ref.~\cite{berg}, we obtain
\begin{equation}
v \sigma ( \widetilde{\chi}^0_1 \widetilde{\chi}^0_1 \to \gamma\gamma)
\simeq 1 \times 10^{-28} \;{\rm cm}^{3} {\rm s}^{-1} \;,
\end{equation}
which is roughly independent of the mass of the neutralino from a few hundred
GeV to a few TeV.  For comparison, the value of $v\sigma$ for a pure wino case
is about $14 \times 10^{-28} \;{\rm cm}^{3} {\rm s}^{-1}$ \cite{cw}.  The
photon flux as a result of this annihilation is given by \cite{berg2}
\begin{eqnarray}
\Phi_\gamma &\simeq& 1.87 \times 10^{-11} \, \left( \frac{N_\gamma v\sigma}
{10^{-29}\,{\rm cm}^3 {\rm s}^{-1}} \right ) \, \left( \frac{10\,{\rm GeV}}
{M_{\widetilde{\chi}^0_1}} \right )^2 \; J(\psi)\;
{\rm cm}^{-2} {\rm s}^{-1} {\rm sr}^{-1} \nonumber \\
&\simeq & 1.5 \times 10^{-11} \;{\rm cm}^{-2} {\rm s}^{-1} {\rm sr}^{-1}  \;,
\end{eqnarray}
where we have used $v\sigma=1\times 10^{-28}\,{\rm cm}^{3} {\rm s}^{-1}$,
${M_{\widetilde{\chi}^0_1}}=500$ GeV, $N_\gamma=2$, and $J(\psi=0)=100$ for the
photon flux coming from the Galactic Center.  The value of $J(\psi)$ depends on
the selected Galactic halo model, which ranges from $O(10)$ to $O(1000)$
\cite{berg2}.  For a typical Atmospheric Cerenkov Telescope (ACT) such as
VERITAS \cite{veritas} and HESS \cite{hess}, the angular coverage is about
$\Delta \Omega =10^{-3}$ and may reach a sensitivity at the level of $(10^{-14}
- 10^{-13}) \, {\rm cm}^{-2}\, {\rm s}^{-1}$ at the TeV scale.  Therefore, the
signal of pure Higgsino dark matter annihilating into monochromatic photons is
easily covered by the next generation ACT experiments.

Since the Higgsino annihilation into the $W^+ W^-$ and $ZZ$ pairs is very
effective, one can also measure the excess in anti-protons and positrons
\cite{anti}, which can be measured in anti-matter search experiments, e.g.,
AMSII \cite{ams}, as well as the neutrino flux from the core of the Sun.

\section{Collider Phenomenology \label{sec:pheno}}

The collider phenomenology in the low-$\mu$ split SUSY scenario is mainly
concerned with the production and detection of neutralinos and charginos.  We
restrict our discussions below to the case with exact or approximate $R$-parity
symmetry.

\subsection{Gaugino pair production}

Note that the masses of the two lightest neutralinos and the lightest chargino
come from a single parameter $\mu$.  They are almost degenerate in mass at tree
level, and the mass splitting is of $O(10)$ keV, as shown in Sec.~\ref{sec:sp}.
The radiative corrections lift the mass degeneracy such that the lightest
chargino is slightly heavier than the lightest neutralino by a mass splitting
at least $340$ MeV.

\begin{figure}[t!]
\centering
\includegraphics[width=5.5in]{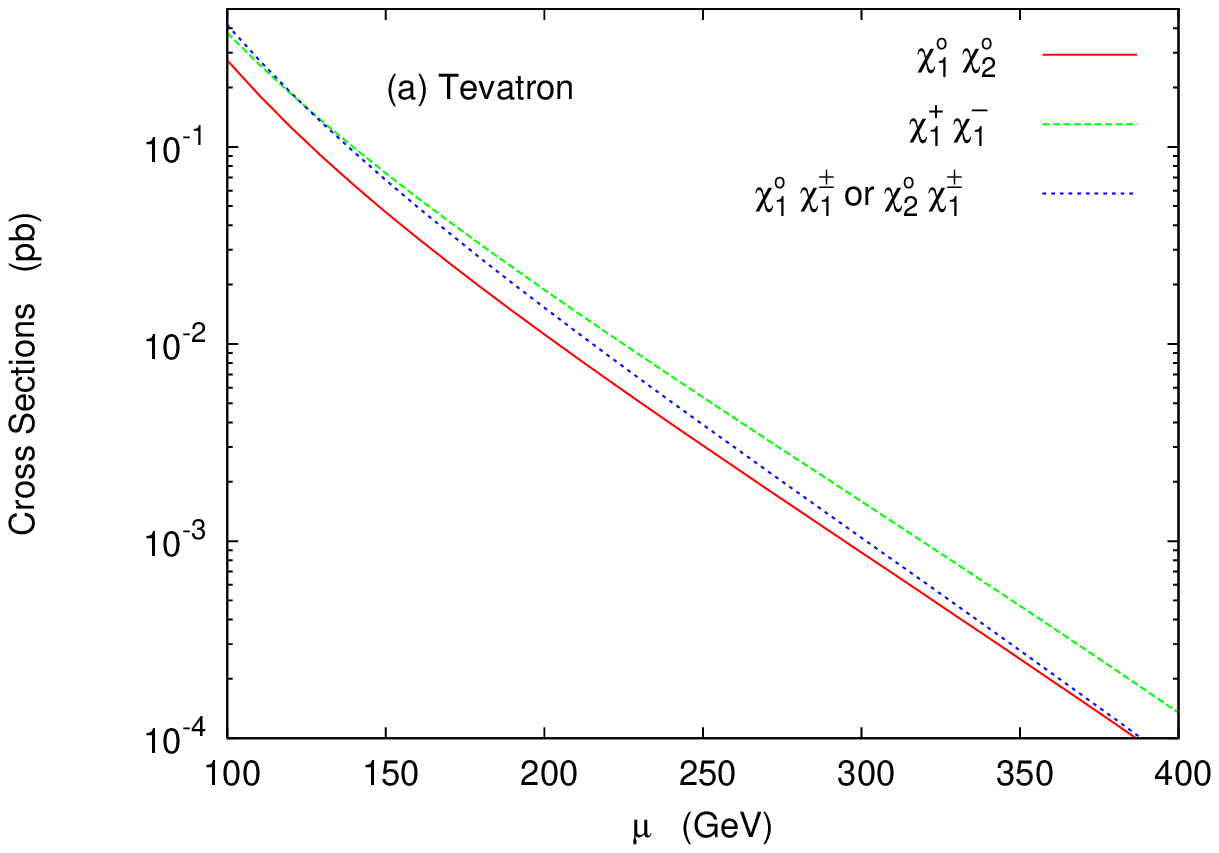}
\includegraphics[width=5.5in]{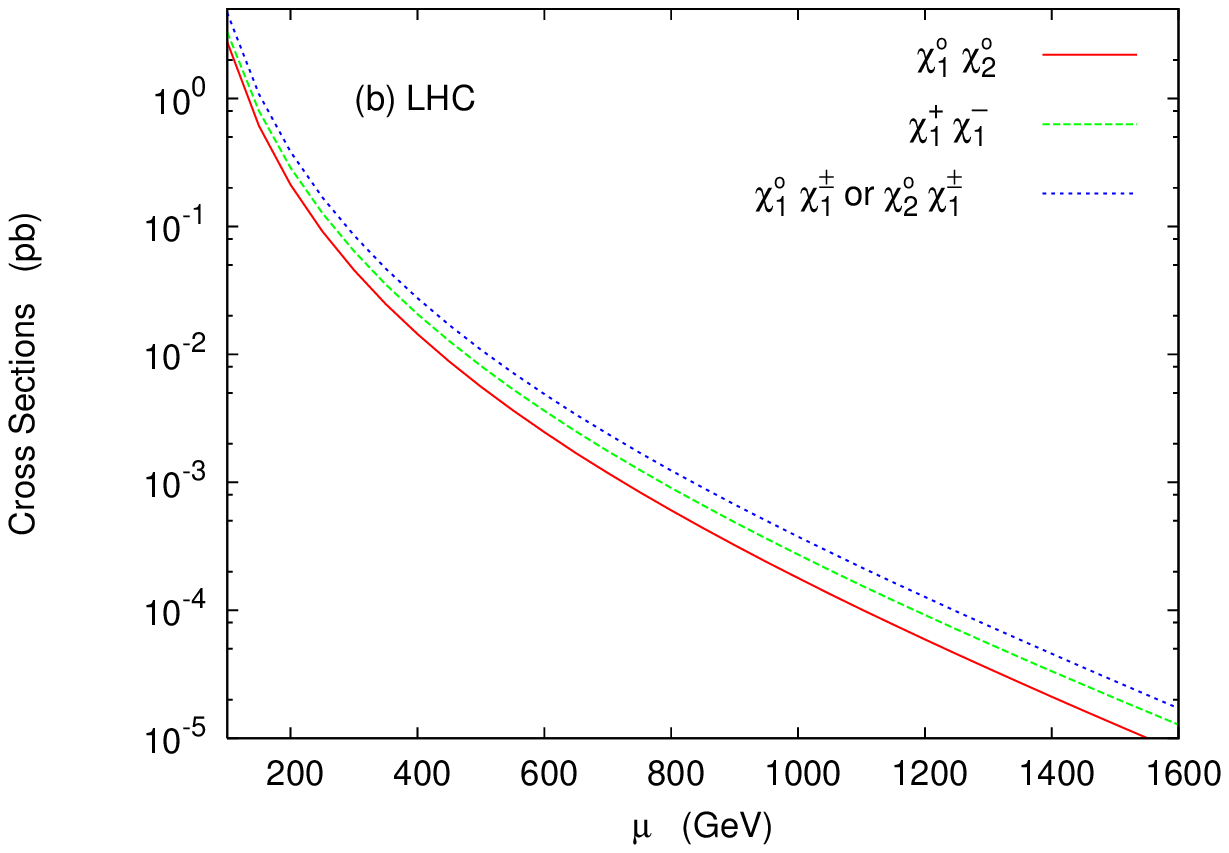}
\caption{\small \label{rate}
Production cross sections versus $\mu$ (the Higgsino mass
parameter at the weak scale) for the
$\widetilde{\chi}^0_1 \widetilde{\chi}^0_2$,
$\widetilde{\chi}^+_1 \widetilde{\chi}^-_1$,
$\widetilde{\chi}^0_1 \widetilde{\chi}^\pm_1$ ($\pm$ states summed), and
$\widetilde{\chi}^0_2 \widetilde{\chi}^\pm_1$ channels at (a) the Tevatron and
(b) the LHC.
}
\end{figure}

The only pair production channels at hadronic colliders are
$\widetilde{\chi}^0_{1} \widetilde{\chi}^0_{2}$, $\widetilde{\chi}^+_1
\widetilde{\chi}^-_1$, and $\widetilde{\chi}^0_{1,2} \widetilde{\chi}^\pm_{1}$
through the Drell-Yan process.  Note that vanishing couplings of
$Z$-$\widetilde{\chi}^0_{1}$-$\widetilde{\chi}^0_{1}$ and
$Z$-$\widetilde{\chi}^0_{2}$-$\widetilde{\chi}^0_{2}$ suppress the production
of $\widetilde{\chi}^0_{1} \widetilde{\chi}^0_{1}$ and $\widetilde{\chi}^0_{2}
\widetilde{\chi}^0_{2}$.  Since the only relevant parameter is $\mu$, we expect
that their cross sections show fixed ratios.  In Fig.~\ref{rate} we present the
production cross sections versus the $\mu$ parameter for the Tevatron and LHC.
We sum the cross sections of $\widetilde{\chi}^0_1 \widetilde{\chi}^+_1$ and
$\widetilde{\chi}^0_1 \widetilde{\chi}^-_1$ in the figure.  Note that at the
Tevatron ($p\bar p$ collision at 1.96 TeV), the production cross sections for
$\widetilde{\chi}^0_1 \widetilde{\chi}^+_1$ and $\widetilde{\chi}^0_1
\widetilde{\chi}^-_1$ are the same.  Furthermore, since $\widetilde{\chi}^0_1$
and $\widetilde{\chi}^0_2$ are almost degenerate in mass, the curves for the
production cross sections of $\widetilde{\chi}^0_1 \widetilde{\chi}^\pm_1$ and
$\widetilde{\chi}^0_2 \widetilde{\chi}^\pm_1$ are indistinguishable in the
figure.

\begin{figure}[t!]
\centering
\includegraphics[width=5.8in]{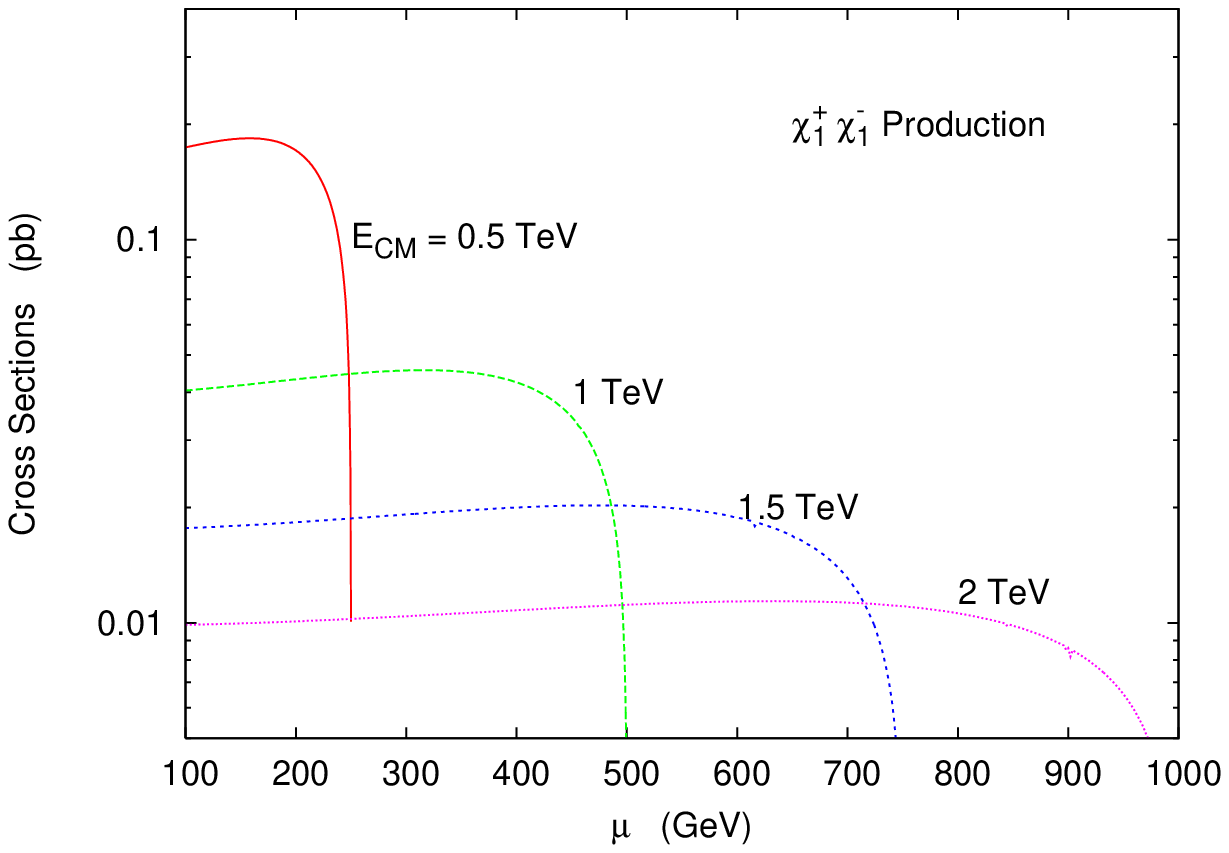}
\includegraphics[width=5.8in]{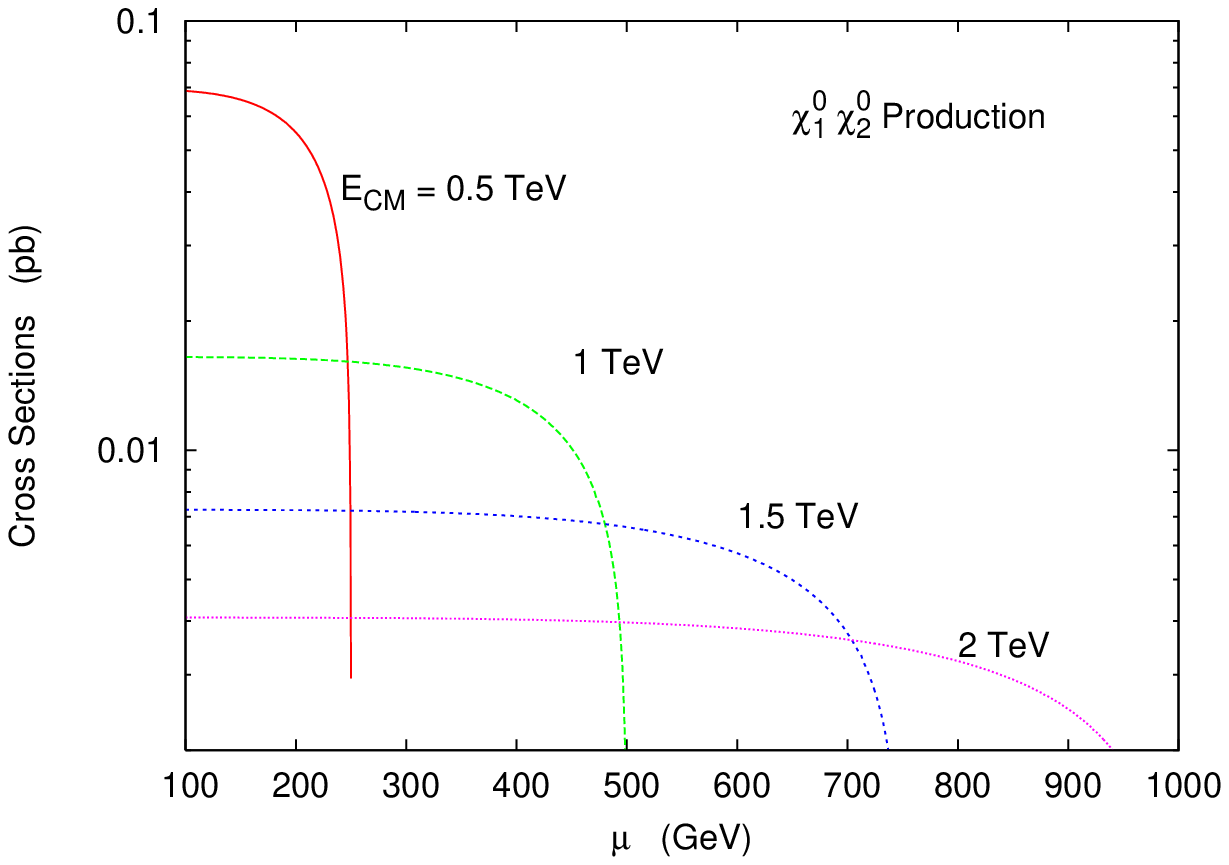}
\caption{\small \label{eerate}
Production cross sections versus $\mu$ (the Higgsino mass
parameter at the weak scale)
for the
$\widetilde{\chi}^+_1 \widetilde{\chi}^-_1$ and
$\widetilde{\chi}^0_{1} \widetilde{\chi}^0_2$
at $e^+ e^-$ linear collider for $\sqrt{s} = 0.5, 1, 1.5, 2$ TeV.
}
\end{figure}

At $e^+ e^-$ linear colliders, one can consider neutralino pair
$\widetilde{\chi}^0_{1} \widetilde{\chi}^0_{2}$ and chargino pair
$\widetilde{\chi}^+_{1} \widetilde{\chi}^-_{1}$ production.  We show their
cross sections in Fig. \ref{eerate}.  Since the $\mu$ parameter is the only
parameter, the cross sections show interesting ratios, given by the gauge
coupling.

\subsection{Neutralino and chargino decays}

In the case of the lightest neutralino as the LSP, the lightest chargino will
decay into the lightest neutralino plus leptons or jets.  Even though the mass
degeneracy is lifted by radiative corrections, the mass difference is quite
small and only of the order of twice the pion mass.  The decay products are
thus very soft.  Experimentally, it is a challenge to tag these soft leptons or
jets.  The decay rate of the chargino into the neutralino and a virtual $W$
boson depends critically on the mass difference $\Delta M_+ \equiv
m_{\widetilde{\chi}^+_1} - m_{\widetilde{\chi}^0_1}$.  The phenomenology in
this case had been studied in great detail in Ref.~\cite{chen}.  We give
highlights as follows.

The partial decay width of $\widetilde{\chi}^+_1 \to \widetilde{\chi}^0_1 f
\bar f'$ is given by \cite{chen}
\begin{eqnarray}
 \Gamma (\widetilde{\chi}^+_1 \to \widetilde{\chi}^0_1 f \bar f')\!\!\!\!&&
\nonumber \\
= \frac{N_c G_F^2}{(2\pi)^3} \biggr\{ \!\!\!&&\!\!\!
M_{\widetilde{\chi}^+_1}
 \left[
   \left( O^L_{11} \right)^2 + \left( O^R_{11} \right)^2 \right] \no
   \\ && \times
 \int_{( M_{\widetilde{\chi}^0_1} + m_f )^2}^{M^2_{\widetilde{\chi}^+_1
}} d q^2 \, \left( 1 - \frac{ M^2_{\widetilde{\chi}^0_1} + m^2_f} {q^2}
           \right ) \,
   \left( 1- \frac{q^2}{ M^2_{ \widetilde{\chi}^+_1} } \right )^2
    \lambda^{1/2}( q^2,   M^2_{\widetilde{\chi}^0_1}, m^2_f )
                \nonumber \\
&& \!\!\!\!
- 2  M_{\widetilde{\chi}^0_1} O^L_{11} O^R_{11}  \,
  \int_{m^2_f}^{ \Delta M_+^2}
 dq^2 \; \frac{q^2}{  M^2_{\widetilde{\chi}^+_1}}
 \left( 1- \frac{ m^2_f}{q^2} \right )^2
  \,     \lambda^{1/2}(M^2_{\widetilde{\chi}^+_1},
 M^2_{\widetilde{\chi}^0_1}, q^2 )  \biggl\} ~,
\label{decay}
\end{eqnarray}
where $(f,f')$ is, for example, $(u,d)$ or $(e,\nu_e)$, $N_c = 3 (1)$ if $f$ is
a quark (lepton), and $\lambda(a,b,c) = (a+b-c)^2 - 4 a b$.  In low $\mu$-split
SUSY, $|O_{11}^L| = |O_{11}^R| = 1/2$.  The above formula is valid for (i)
leptonic decays and (ii) hadronic decays when $\Delta M_+ \agt 2$ GeV.  For
hadronic decays with $\Delta M_+ \alt 1-2$ GeV, one has to explicitly sum over
exclusive hadronic final states.  We have to include the partial decay widths
for the decays into one, two, and three pions.  The explicit formulas can be
found in Ref.~\cite{chen}.

Generally speaking, the detection of the chargino depends on the size of
$\Delta M_+$:

\begin{enumerate}
\item $\Delta M_+ < m_\pi$.  As we have explained before, this case will not
  happen due to the one-loop radiative corrections that result in a mass
  splitting greater than the pion mass.
\item $m_\pi < \Delta M_+ < 1$ GeV.  This is the most difficult regime to probe
  experimentally, and very much depends on the design of the central detector.
  Important criteria are the decay length $c\tau$ of the chargino and the
  momentum of the pion from the chargino decay.  The decay length $c\tau$ may
  be long enough for the chargino to travel through a few layers of the silicon
  vertex detector.  For example, if $m_\pi < \Delta M_+ < 190$ MeV the chargino
  will typically pass through at least two layers of silicon chips \cite{chen}.
  Since the pion is produced from the chargino decay, it is a non-pointing
  pion; the backward extrapolation of the pion track does not lead to the
  interaction point.  The resolution on the impact parameter $b_{\rm res}$
  depends on the momentum of the pion $p_\pi \sim \sqrt{ \Delta M^2_+ -
    m^2_\pi}$ in the chargino rest frame.  The higher the momentum is, the
  better the resolution $b_{\rm res}$ will be \cite{chen}.  Thus, detecting the
  signal involves the combination of detecting a track left in only a few
  layers of the silicon detectors and identifying a nonzero impact parameter of
  the pion coming out of the chargino.  A detailed simulation is beyond the
  scope of the present paper.
\item $\Delta M_+ \agt 1-2$ GeV.  We can use Eq.~(\ref{decay}) to estimate the
  total decay width of the chargino.  The decay width is large enough that the
  decay is prompt, producing soft leptons, pions, or jets, plus large missing
  energy.  The problem is on the softness of the decay products, whose
  detection is experimentally difficult.  Only when $\Delta M_+$ is
  sufficiently large to produce hard enough leptons or jets can the chargino
  decay be detected.  Otherwise, one has to rely on some other channels, such
  as $e^+ e^-\to \gamma \widetilde{\chi}^+_1 \widetilde{\chi}^-_1 \to \gamma +
  \rlap/{E}_T$, a single photon plus large missing energy above the SM
  background $e^+ e^- \to \gamma \nu\bar \nu$ \cite{chen}.  Unfortunately, the
  signal rate is ${\cal O}(\alpha_{\rm em})$ smaller than the chargino pair
  production.  Detecting such a signal is even more difficult at hadronic
  colliders.
\end{enumerate}

In summary, the detection of the chargino is easier only when $\Delta M_+$ is
larger than a few GeV.  As we show in Sec.~\ref{sec:sp}, $\Delta M_+$ from the
one-loop corrections is at least 340 MeV.  The intermediate range between this
value and 1 GeV poses a challenge to experiments.  The questions are how many
layers of silicon the chargino can travel and how well the resolution of the
non-pointing pion can be.

On the other hand, the mass splitting between the second lightest and the
lightest neutralino is even smaller than the electron mass when $\tilde{m} \agt
10^{8-9}$ GeV.  Only $\widetilde{\chi}^0_2 \to \widetilde{\chi}^0_1 \, \gamma$
and $\widetilde{\chi}^0_2 \to \widetilde{\chi}^0_1 \nu \bar{\nu}$ are possible.
The former decay mode produces a soft photon with keV energy while the latter
leads to a totally missing energy signal.  In addition, the decay length is of
the order of $10^9 - 10^{12}$ m for $\tilde{m} =10^8 - 10^9$ GeV.  Therefore,
it is impossible to detect this second lightest neutralino as long as the mass
splitting $\Delta M_{12}$ is less than twice the electron mass.  We will have
an excess of the totally missing energy signal over the SM background of $e^+
e^- \to \nu \bar{\nu}$ at $e^+ e^-$ colliders.  In the case of $\tilde{m} <
10^7$ GeV, the mass splitting $\Delta M_{12}$ is larger than the electron mass.
The charged lepton mode is open but the leptons are still too soft for
identification.

In this almost hopeless scenario, one has to rely on additional tags of the
processes, such as $e^+ e^- \to \widetilde{\chi}^+_1 \widetilde{\chi}^-_1
\gamma$ at $e^+ e^-$ colliders \cite{drees} or $q \bar q \to
\widetilde{\chi}^+_1 \widetilde{\chi}^-_1 g\,(\gamma),\; \widetilde{\chi}^+_1
\widetilde{\chi}^0_{1,2} g\,(\gamma)$ at hadron colliders.

\section{Conclusions \label{sec:summary}}

In the present paper we have considered a low-$\mu$ split SUSY scenario.  The
only parameter beyond the SM is the Higgsino mass parameter $\mu$, which gives
rise to two light neutralinos and a pair of light charginos.  We summarize its
characteristic features as follows:

\begin{enumerate}
\item Gauge coupling unification is as good as that in MSSM or split SUSY.  The
  unification point is lower at $10^{14-15}$ GeV.
  
\item The neutral Higgsino is the LSP.  The mass degeneracy among the two
  lightest neutralinos and the lightest chargino gives a very large
  coannihilation effect such that the thermal source of Higgsino dark matter
  dominates only if the Higgsino mass is at least 1 TeV.  On the other hand, it
  can have other non-thermal sources, which involve other unknown parameters of
  the model.
  
\item The Higgsino dark matter has negligible direct detection rates, which
  only arise from loop corrections.  However, the pair annihilation cross
  sections into $WW,ZZ,\gamma\gamma$ are large.  Thus, we can look for positron
  or anti-proton excess in nearby galaxies, as well as monochromatic photon
  lines from the Galactic Center.
  
\item The collider phenomenology is also quite different from the usual MSSM or
  split SUSY.  The only extra parameter beyond the SM is the $\mu$ parameter.
  The electroweak gaugino pair production shows a fixed ratio because they are
  produced via the gauge couplings.
  
\item The decays of the second lightest neutralino and the lightest chargino
  are difficult to detect because the chargino decay length is not long enough
  and its decay products are too soft.  One has to rely on additional tags of
  the processes.

\end{enumerate}

\section*{Acknowledgments}

K.C. thanks E.J. Chun and Y. Nomura for discussions during a YITP workshop and
the hospitality of Nobuchika Okada at the KEK.  C.-W.~C.\ is grateful to the
hospitality of the National Center for Theoretical Sciences, Taiwan, during his
visit when part of the work was done.  This research was supported in part by
the National Science Council of Taiwan R.~O.~C.\ under Grant Nos.\ NSC
94-2112-M-007-010- and NSC 94-2112-M-008-023-.  The work of J.S. was supported
by the Korea Research Foundation Grant (KRF-2005-070-c00030).

{\it Note added}: During the write-up a couple of works
\cite{Vereshkov,Mahbubani} similar to the present one appears.  We have
different discussions with the former paper.  We have small overlaps with the
second one.


\end{document}